\documentclass[aps,physrev,preprint]{revtex4-2}
\usepackage{lineno}
\usepackage{epsfig,latexsym,cancel,amssymb,amsmath}
\usepackage{graphicx,subfigure}
\usepackage{epstopdf}
\usepackage{hyperref, graphicx, slashed, xcolor, bm}
\usepackage{slashed}

\newcommand{\dd}{{\rm d}}
\newcommand{\bx}{{\bf x}}
\newcommand{\bB}{{\bf B}}
\newcommand{\bE}{{\bf E}}
\newcommand{\bA}{{\bf A}}
\newcommand{\bv}{{\bf v}}
\newcommand{\bJ}{{\bf J}}

\newcommand{\bV}{{\bf V}}
\newcommand{\bR}{{\bf R}}

\newcommand{\dert}{\partial_t}

\begin{document}
\preprint{APS/123-QED}
\title{Solar Reflected Dark Matter under the Influence of a Dark Magnetic Field}
\author{Haoming Nie}
\affiliation{Department of Physics, Tsinghua University, Beijing 100084, China}
\email{nhm20@mails.tsinghua.edu.cn}
\date{\today}
\begin{abstract}
    The scattering of dark matter particles within the Sun’s hot plasma can lead to the acceleration of dark matter, producing a high-energy solar-reflected DM flux detectable in ground-based experiments. In the vector portal model, the dark matter has a sub-MeV-scale mass, and interactions between the dark matter and Standard Model particles are mediated by a hidden vector field--referred to as a dark photon--which kinetically mixes with the conventional photon through a small mixing angle. Furthermore, the solar plasma generates intense magnetic fields. Due to the photon-dark photon mixing, this simultaneously sources a ``dark magnetic field". For sufficiently low dark photon masses, this dark magnetic field is capable of deflecting dark matter particles traversing the Sun. We found that if the dark magnetic force is sufficiently strong, the dark magnetic field becomes a wall, preventing the dark matter particles from reaching the deep core region, suppressing their reflected flux. This scenario corrects the sensitivity of the solar-reflected dark matter detection, offering critical insights for ground-based experiments aiming to probe dark matter.
\end{abstract}

\maketitle

\section{Introduction}
\label{sec:intro}

The Standard Model (SM) is a well-established theory for describing the Universe. However, there are still many reasons for us to extend this structure. One major front of beyond the Standard Model physics is dark matter (DM). From astrophysical observations, we realize that a large portion of matter in the Universe cannot be explained by the Standard Model. To reveal the nature of DM, various direct detection experiments are set up to capture the feeble signal of DM-SM interactions. Currently, the experimental exclusion line of the galactic halo DM is reaching the threshold of elastic scattering of the neutrino background (the ``neutrino fog")~\cite{XENON:2024ijk,PandaX:2024muv}. Therefore, there is increasing interest in alternative DM models other than the traditional weakly interacting massive particles paradigm, especially MeV- or sub-MeV-scale light DM~\cite{SENSEI:2023zdf, DAMIC-M:2023gxo, SuperCDMS:2023sql, XENON:2017vdw, XENON:2022ltv, LUX:2020car, LZ:2023poo, PandaX:2022xqx, CDEX:2022kcd, CDEX:2022dda}. For light DM with an MeV or sub-MeV mass, the scattering between nuclei is inefficient due to a large mass hierarchy. Therefore, detection through an electron recoil signal is a better choice. However, the recoil energy is still small and often falls below the detection thresholds of experiments, making direct detection of light DM a tough job, especially when the mass of DM is even smaller than an MeV, as is the case considered in this paper.

In the last decade, it has been proposed in Ref.~\cite{An:2017ojc} that DM particles can be accelerated via elastic scatterings with keV-energy electrons in the hot, dense plasma of the solar interior. Since the masses of the electron and sub-MeV-scale light DM particles are of the same scale, the energy transfer is efficient, producing a solar reflected DM (SRDM) spectrum with a significant keV-energy tail. Taking into account the SRDM flux, exclusion limits derived from direct detection experiments are established in Refs.~\cite{An:2017ojc,An:2021qdl,Emken:2021lgc,Emken:2024nox}. The constraints of cross sections are around $\sigma_e\sim10^{-38}\text{cm}^2$ for contact interactions, and constraints of millicharges are around $Q_{\text{eff}}\sim 10^{-9}$ for interactions mediated by a massless dark photon.

On the theoretical side, direct detection of both halo DM and SRDM requires a coupling (``portal") between DM and SM particles. Currently we know of three portals to connect the DM and the Standard Model: the neutrino portal, the Higgs portal, and the vector portal. The vector portal involves an additional vector field $V^{\mu}$, often called the dark photon~\cite{1982DP,Holdom:1985ag,HOLDOM198665,HOLDOM1991329,Dienes:1996zr}, that mixes with the $\mathrm{U}(1)_Y$ hypercharge field by a mixing angle denoted by $\kappa$. After electroweak symmetry breaking this provides a mixing between the dark photon and the normal photon. Recently, the study of dark photons has become a prosperous front for both theorists and experimentalists. Dark photons with masses larger than 10 MeV are subjected to traditional collider search, while light dark photons are more limited by astrophysical and cosmological bounds. A thorough illustration of dark photon constraints can be found in Ref.~\cite{Caputo:2021eaa}.

In vector portal models, the dark sector typically includes at least two particles: a dark photon and a heavy DM with a dark photon U(1) charge. The scattering of a DM particle off electrons is mediated by the exchange of a dark photon-photon mixed state. When the dark photon mass significantly exceeds the characteristic momenta, $m_V\gg p_e,p_\chi$, the interaction reduces to an effective contact interaction. Conversely, if $m_V\ll p_e,p_\chi$, then the scattering is well approximated by massless mediator scattering. In previous studies, the mass of the dark photon appeared solely as a parameter in the propagator within the scattering amplitude. For $m_V$ much smaller than the temperature of the Sun ($\sim\text{keV}$), the dark photon is treated as massless and any $m_V$ gives the same result. However, in this paper, we point out that DM-SM interactions represent only one aspect of the dark photon's phenomenology. When the mass of the dark photon corresponds to an astronomical scale, $m_V\sim R_\odot^{-1}$, where $R_\odot$ is the radius of the Sun, the large-scale dark photon field significantly influences the motion of SRDM. The Sun sustains a strong Standard Model magnetic field, generated by the electric current induced by the dynamical plasma inside. Due to the dark photon-photon mixing, such an electric current simultaneously sources a ``dark magnetic field". Any DM particle approaching the Sun will be affected by a dark magnetic force with a strength roughly equivalent to $\sim \kappa e_D v |\nabla^2 \bB|/m_V^2$, where $e_D$ is the ``dark charge" carried by DM and $v$ is the velocity of the DM particle. For $m_V\lesssim10^{-14}\text{eV}$ and $\kappa e_D>10^{-9}$, the dark magnetic force is large enough to significantly redirect DM inside the Sun, giving rise to a new phenomenology of SRDM. In our simulations, the effect of the dark magnetic field is double-edged. The dark magnetic force will bend the trajectory of the DM particles, causing them to stay longer inside the Sun, potentially inducing more scattering. On the other hand, the deflection prevents DM particles from reaching the core region of the Sun, thus suppressing the scattering probability with energy transfer $\gtrsim1\,\text{keV}$. Our simulations reveal that the high-energy tail of SRDM is suppressed, while the $\sim 10\,\text{eV}$ part of the spectrum is enhanced. By taking the dark magnetic force into account, we can draw more appropriate lines of constraints on the millicharge $Q_{\text{eff}}$ for different $m_V$.

This paper is organized as follows. In Sec.~\ref{sec:eom}, we derive the formula of dark electric and magnetic field induced by electric charge and current. In Sec.~\ref{sec:solarDMF}, we build a model of dark magnetic field generated by solar dynamics, and show the behavior of DM particle in a dark magnetic field. In Sec.~\ref{sec:SRDM}, we briefly explain how SRDM flux is generated by simulations. In Sec.~\ref{sec:detection}, we calculate the signal induced by SRDM in the detection target of experiments and show the exclusion line with the correction of a dark magnetic field. Then in the last section, we give a summary of this paper.

\begin{figure}
    \centering
    \includegraphics[width=0.7\linewidth]{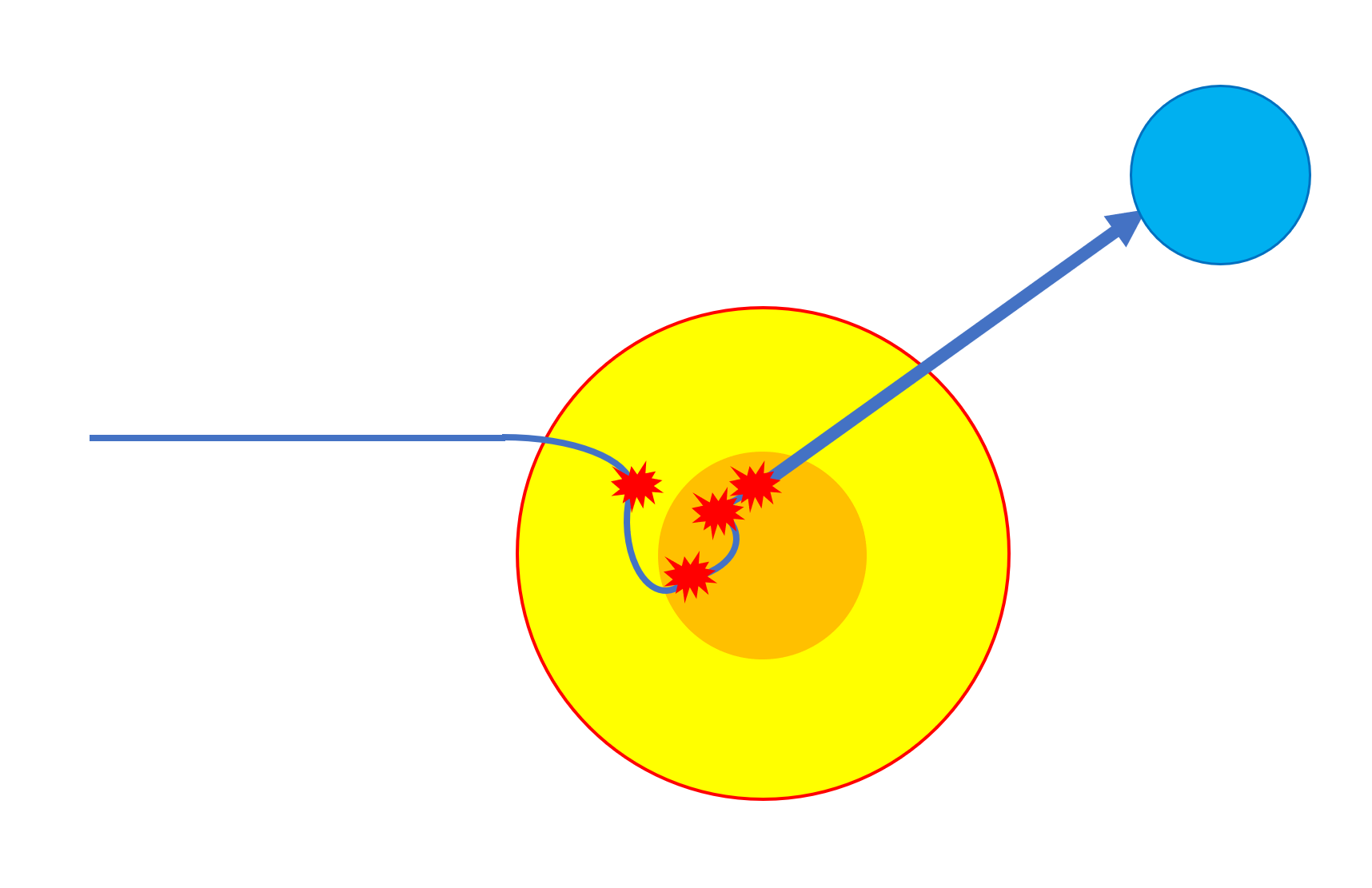}
    \caption{Illustration of the basic ideas of solar reflected dark matter.}
    \label{fig:SRDM}
\end{figure}

\section{Dark Photon Equations of Motion}
\label{sec:eom}
In our dark sector model, the dark sector is composed of a DM particle with a sub-MeV mass, labeled $\chi$, charged under a dark U(1) charge and a dark photon. The DM with sub-MeV mass is the main contribution to the DM abundance, and the interactions between $\chi$ and the SM particles are mediated by the mixed state of dark photon and photon.

After electroweak symmetry breaking, the dark photon model can be described as an additional vector field with a Stuckelberg mass term, mixed with the usual electromagnetic field with a mixing angle $\kappa$. The Lagrangian can be written as
\begin{equation}
    \mathcal{L}=-\frac{1}{4}F_{\mu \nu }^2-\frac{1}{2} \kappa  F_{\mu \nu } V^{\mu \nu } -\frac{1}{4}V_{\mu \nu }^2 + \frac{1}{2} m_V^2 V^{\mu } V_{\mu } - e_D J_D^{\mu } V_{\mu } - e J^{\mu } V_{\mu },
\end{equation}
where $V_\mu$ is the dark photon field, $V^{\mu\nu}=\partial^\mu V^\nu-\partial^\nu V^\mu$ is the dark photon field strength, $J_D^\mu=(\rho_D,\bJ_D)$ is the dark 4-current, $\rho_D$ is the dark charge density, and $\bJ_D$ is the dark charge current. The classical equations of motion can easily be derived through the Lagrangian equation:
\begin{align}
    \partial_\mu \left(F^{\mu \nu }+\kappa  V^{\mu \nu }\right)&=e J^{\nu }\\
    \partial_\mu \left(\kappa  F^{\mu \nu }+V^{\mu \nu }\right)+m_V^2 V^{\nu }&=e_D J_D^{\nu }.
\end{align}
Writing in 3D components, the equations of motion are
\begin{align}
    -\dert\tilde{\bE}+\nabla \times \tilde{\bB}-\kappa \dert\bE+\kappa \nabla \times \bB+m_V^2 \bV&=e_D \bJ_D\label{DarkEM:1}\\
    -\dert\bE+\nabla \times \bB-\kappa  \dert\tilde{\bE}+\kappa \nabla \times \tilde{\bB}&=e \bJ\label{DarkEM:2}\\
    \nabla \cdot \tilde{\bE}+\kappa \nabla \cdot \bE+m_V^2 V^0&=e_D \rho _D\label{DarkEM:3}\\
    \nabla \cdot \bE+\kappa \nabla \cdot \tilde{\bE}&=e \rho\label{DarkEM:4},
\end{align}
where $\tilde{\bE}=-\nabla V^0-\dert\bV$ and $\tilde{\bB}=\nabla\times\bV$ are the dark electric and magnetic fields. For a massive vector field, we must have $\dert V^0+\nabla\cdot\bV=0$. Taking the curl of Eq.~\eqref{DarkEM:1}, assuming static fields and neglecting dark charges, we have
\begin{equation}
    m_V^2\tilde{\bB}-\nabla^2\tilde{\bB}=\kappa\nabla^2\bB.
\end{equation}
Therefore, for $m_V$ not too much smaller than $1/R_B$, where $R_B$ is the spatial variation scale of $\bB$, we can approximately write $\tilde{\bB}\sim\kappa/m_V^2\times\nabla^2\bB\sim\kappa B/m_V^2 R_B^2$.

In terms of scalar and vector potentials, Eqs.~\eqref{DarkEM:1}--\eqref{DarkEM:4} can be rewritten as
\begin{align}
    \dert^2 V^0-\nabla ^2V^0+\frac{m_V^2}{1-\kappa^2}V^0&=\frac{e_D}{1-\kappa ^2}\rho_D-\frac{\kappa e}{1-\kappa^2}\rho\\
    \dert^2 \bV-\nabla ^2\bV+\frac{m_V^2}{1-\kappa^2}\bV&=\frac{e_D}{1-\kappa^2}\bJ_D-\frac{\kappa e}{1-\kappa^2}\bJ.
\end{align}
Here, it is clear that the normal electric charge and current are also sources of dark electric and magnetic fields, but the coupling is suppressed by a millicharge $Q_{\text{eff}}=\kappa e_D/e$. In the same virtue, we can also derive the equations for the normal potentials:
\begin{align}
    \partial_t^2\phi-\nabla ^2\phi &=\frac{e}{1-\kappa^2}\rho-\frac{\kappa e_D}{1-\kappa^2}\rho _D-\frac{\kappa m_V^2}{1-\kappa^2}V^0\label{eq:epot}\\
    \partial_t^2\bA-\nabla^2\bA&=\frac{e}{1-\kappa^2}\bJ-\frac{\kappa e_D}{1-\kappa^2}\bJ_D-\frac{\kappa m_V^2}{1-\kappa ^2}\bV\label{eq:mvpot}.
\end{align}
The dark photon field enters as a source for the normal electromagnetic potential. However, because of the smallness of $\kappa$, the influence on the usual electromagnetic field is always negligible in real situations where DM is rare, $\rho_D,\bJ_D\approx0$: When $m_V\gg 1/R_B$, $V^0\approx-(\kappa e/m_V^2)\rho$, $\bV\approx-(\kappa e/m_V^2)\bJ$. Therefore, the third terms on the rhs in Eqs.~\eqref{eq:epot} and \eqref{eq:mvpot} are $\kappa^2 e\rho$ and $\kappa^2 e\bJ$, which are much smaller than the first terms on the rhs. Conversely, when $m_V\ll 1/R_B$, the equations of $V^\mu$ become the same as $A^\mu$, except for the fact that the dark photon is coupled with the normal electric charge by $\kappa e$. Hence $V^0\approx\kappa\phi$ and $\bV\approx\kappa\bA$. The third terms on the rhs in Eqs.~\eqref{eq:epot} and \eqref{eq:mvpot} are $\kappa^2 m_V^2\phi$ and $\kappa^2 m_V^2\bA$, which are always much smaller than the terms on the lhs.

In reality, DM particles are very few, $\rho_D\approx0$, $\bJ_D\approx0$. For qualitative estimation, we only consider a static field, then
\begin{align}
    V^0-\frac{1-\kappa^2}{m_V^2}\nabla^2 V^0&=-\frac{\kappa e}{m_V^2}\rho\\
    \bV-\frac{1-\kappa ^2}{m_V^2}\nabla ^2\bV&=-\frac{\kappa e}{m_V^2}\bJ.
\end{align}
These are standard massive boson equations of motion with a source. The solution should be the famous Yukawa potential:
\begin{align}
    V^0(\bx)&=-\frac{\kappa}{4\pi (1-\kappa^2)}\int d^3 x'\,e\rho(\bx')\frac{e^{-(m_V/\sqrt{1-\kappa ^2})\left|\bx-\bx'\right|}}{\left|\bx-\bx'\right|}\label{Yukawa1}\\
    \bV(\bx)&=-\frac{\kappa}{4\pi (1-\kappa^2)}\int d^3 x'\,e\bJ(\bx')\frac{e^{-(m_V/\sqrt{1-\kappa ^2})\left|\bx-\bx'\right|}}{\left|\bx-\bx'\right|}\label{Yukawa2}.
\end{align}
Because of the decaying exponential factors, the integrals are usually hard to perform analytically. In this paper, we calculate them by numerical integration. After $\bV$ is calculated, we can derive $\tilde{\bB}=\nabla\times\bV$ directly.

In the presence of the dark photon field, a DM particle with velocity $\bv$ will experience a force just like electrons in the normal electromagnetic field,
\begin{equation}
    \mathbf{F}_{\text{dark}}=e_D(\tilde{\bE}+\bv\times\tilde{\bB}).
\end{equation}
Therefore, we can make an estimation of the force strength, $F_{\text{dark}}\sim\kappa e_Dv B/m_V^2R_B^2$. The cyclotron radius corresponding to this ``dark Lorentz force" is
\begin{equation}
    R=\frac{m_\chi m_V^2 R_B^2 v}{\kappa e_D B},\label{DLF}
\end{equation}
where $m_\chi$ is the mass of DM $\chi$. If we want the effect of this force to be significant, this radius should not be larger than the radius of the Sun. Consequently, we derive an estimation for the parameters:
\begin{equation}
    \frac{m_\chi m_V^2}{\kappa e_D}\lesssim B \frac{R_\odot}{R_B^2 v}.
\end{equation}
The current direct detection constraint is $\kappa e_D\sim10^{-10}$. We choose the benchmark parameters $m_\chi\sim 0.1\,\text{MeV}$, $v\sim0.002$, $R_B\sim0.04R_\odot$, $B\sim 0.4\,\text{G}$; the order-of-magnitude estimate for $m_V$ is
\begin{equation}
    m_V\lesssim 10^{-14}\,\text{eV}.
\end{equation}
In actual simulations (see Sec.~\ref{sec:SRDM}) we found that around $m_V\sim 10^{-14}\,\text{eV}$, there are significant changes in the trajectories if $\kappa e_D\sim 10^{-9}$.

\section{Solar Dark Magnetic Field}
\label{sec:solarDMF}

It is commonly believed that the magnetic field of the Sun is generated by the solar dynamo mechanism in the interior~\cite{Parker1955a,Parker1955b}. Traditionally, this is suspected to be closely related to certain plasma motion patterns in a thin layer called the tachocline (thickness $\sim 0.04 R_\odot$) with a radius around $R_{\text{tac}}\approx0.7R_\odot$~\cite{Strugarek2011tacho,Charbonneau2014}. Recently, new studies suggest that the solar dynamo may originate just below the surface~\cite{vasil2024solar}. The detailed mechanism of magnetic field generation is beyond the scope of this paper. To highlight the main idea of dark magnetic field deflection and give a qualitative estimation, we neglect the small structure eddies and extreme values. We begin with a simple model, assuming there is a thin layer of electric current circling around the Sun in the tachocline region:
\begin{equation}
    e\bJ_{\text{SD}}(r,\theta,\phi)=\frac{3\mathfrak{m}}{4\pi R_{\text{tac}}^3}\sin\theta\frac{1}{\sqrt{2\pi}\sigma_{\text{tac}}}e^{-\frac{(r-R_{\text{tac}})^2}{2\sigma_{\text{tac}}^2}}\hat{\phi},
\end{equation}
Here we choose the Gaussian thickness of the current layer to be $\sigma_{\text{tac}}=0.02R_\odot$. This will generate a dipole magnetic field outside the tachocline:
\begin{equation}
    \bB(r,\theta,\phi)=\frac{\mathfrak{m}}{4\pi r^3}(2\cos\theta\,\hat{r}+\sin\theta\,\hat{\theta}).
\end{equation}
The Sun's surface magnetic field strength may vary according to solar activities and small-scale features on the surface. In some situations, the magnetic field in the photosphere can be as large as kG~\cite{Velez:2008wu}. The author of Ref.~\cite{HighMagneticField} suggests that helioseismology evidence indicates thin layers of $\sim$kG magnetic fields near the surface. However, an extremely high magnetic field exists only at a specific place and/or at a specific time. Since we are considering a large-scale process and modeling the solar magnetic field as a dipole field, the only relevant value is the Solar Mean Magnetic Field (SMMF). We assume that the magnetic field fits the observed SMMF average value on the surface. The value of SMMF varies in different estimations~\cite{howard1974studies,SMMF2018,SMMF}, and we choose the value $0.4\,\text{G}$ from Ref.~\cite{howard1974studies} in this paper. Therefore, $2\mathfrak{m}/4\pi R_\odot^3\sim 0.4\,\text{G}$. This seems to be a strong assumption, but since the DM particles come from all directions, the averaging effect will cause the reflected flux to be irrelevant to the angular distribution of the dark magnetic field, and only the average magnitude at different radii is important. Therefore, this simple model is enough to illustrate the idea qualitatively.

\begin{figure}[htbp]
\centering
\includegraphics[height=0.3\textwidth]{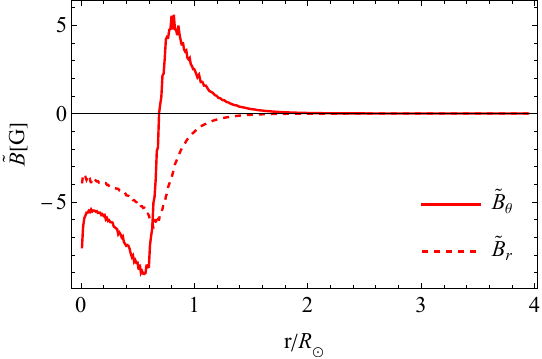}
\qquad
\includegraphics[height=0.3\textwidth]{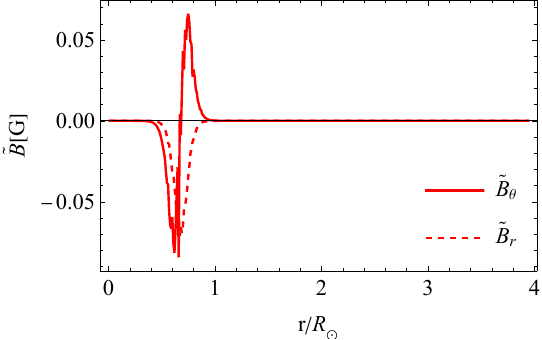}
\caption{Numerically calculated $\tilde{\bB}$, decomposed into radial and azimuthal angle parts: $\tilde{B}_r$ and $\tilde{B}_\theta$. The left graph has $m_V=1\times10^{-15}\,\text{eV}$, while the right graph has $m_V=1\times10^{-14}\,\text{eV}$. The azimuthal angle coordinate is set to be $\theta = 7\pi/10$ in both graphs.\label{fig:VBPlot}}
\end{figure}

The dark photon field $\bV$ profile can be numerically calculated via Eq.~\eqref{Yukawa2}; the resultant dark magnetic field $\tilde{\bB}=\nabla\times\bV=\tilde{B}_r\hat{r}+\tilde{B}_\theta\hat{\theta}$ is shown in Fig.~\ref{fig:VBPlot}. It can be clearly seen that as $m_V$ becomes smaller, the magnitude of the dark magnetic field becomes larger, the distribution in space becomes extensive and approaches a dipole field like the normal magnetic field.

\begin{figure}
    \centering
    \includegraphics[width=0.4\linewidth]{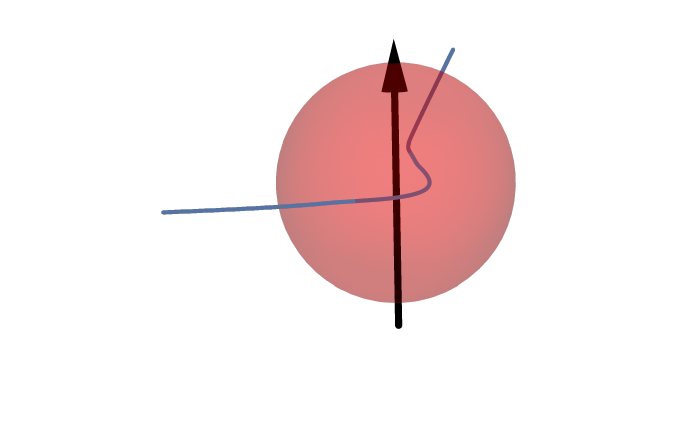}
    \qquad
    \includegraphics[width=0.45\linewidth]{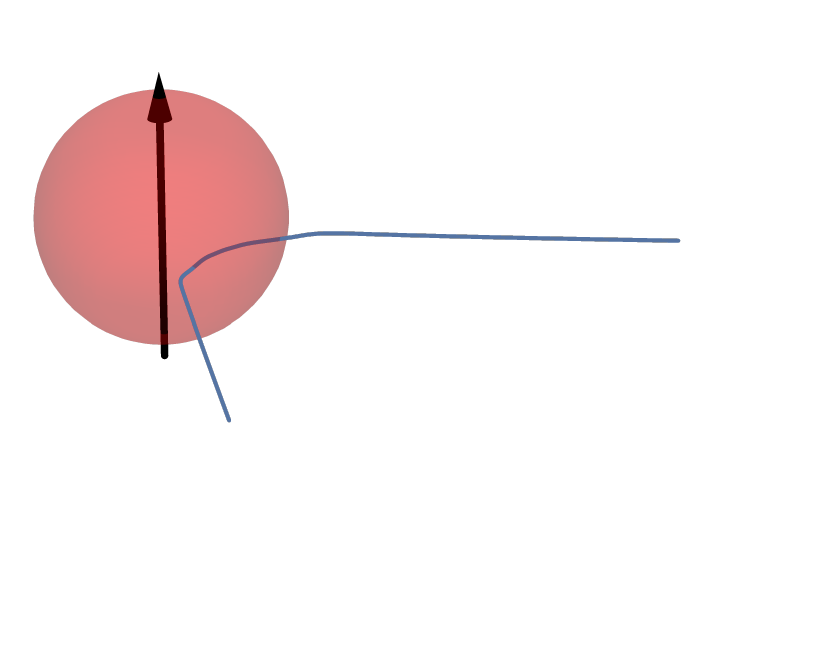}
    \caption{Two illustrative trajectories of DM in the Sun. Both have impact parameter $b=0.3\, R_\odot$, $m_V=1\times 10^{-14}\,\text{eV}$, $m_\chi=0.25\,\text{MeV}$, geometric parameters at infinity $\theta_\infty=\pi/2,\,\nu_\infty=0$ (see Sec.~\ref{sec:SRDM}), and initial velocity $v=220\,\text{km/s}$. The left trajectory has $Q_{\text{eff}}=1.65\times 10^{-8}$, the right one has $Q_{\text{eff}}=3.3\times 10^{-8}$. The black arrow indicates the direction of magnetic moment of the solar dipole field.}
    \label{fig:traj}
\end{figure}

Before conducting simulations, we anticipate that the influence of the dark magnetic field on SRDM will exhibit a dual nature. Firstly, the dark magnetic force induces curvature in the trajectories of DM particles, causing them to spiral within the Sun. This extended path length increases the probability of scattering with solar electrons. Conversely, if the incident angle between an incoming DM particle's velocity vector and the local dark magnetic field line exceeds a critical value, the particle will be deflected outward, preventing its penetration into the hot, dense core region. These competing effects complicate the estimation of the resultant flux spectrum. Generally, DM particles with higher initial velocities possess a higher likelihood of reaching the solar core. The trajectories simulated by our program are illustrated in Fig.~\ref{fig:traj}. Consequently, in the first, for moderate values of kinetic mixing and $m_V\lesssim 10^{-14}\,\text{eV}$, we expect that the low-energy part of the SRDM spectrum is suppressed while the high-energy tail is enhanced. However, simulations in Sec.~\ref{sec:SRDM} indicate the opposite story.

\subsection{Deflection by Dark Magnetic Field}
In order to discuss the deflection of DM by the dark magnetic field qualitatively, we imitate the procedure in plasma dynamics: divide the motion of DM into the motion of the guiding center and the gyromotion of the particle:
\begin{equation}
    \mathbf{r}(t)=\bR(t)+\mathbf{r}_L(t).
\end{equation}
This approximation requires the dark magnetic field to be weakly inhomogeneous slowly changing in a gyroradius $R_{LD}=m_\chi v_\perp/e_D \tilde{B}$. In our estimation, this is not always satisfied by the solar dark magnetic field, but the analysis can still be fruitful for intuition. By expanding everything with $\mathbf{r}_L$ around $\bR$, we arrive at an equation for the guiding center:
\begin{equation}
    \frac{e_D}{m_\chi}\langle\dot{\mathbf{r}}_L\times[\mathbf{r}_L\cdot\nabla \tilde{\bB}(\bR)]\rangle=-\frac{\mathcal{M}_D}{m_\chi}\nabla\tilde{B},
\end{equation}
where $\mathcal{M}_D=\pi R_{LD}^2 J_D$ is the dark magnetic moment of the gyromotion. This derivation can be found in most textbooks on plasma physics, for example, Ref.~\cite{somov2012plasma}. Equivalently, the guiding center experiences a ``force" proportional to the gradient of the dark magnetic field $\mathcal{F}_D=-\mathcal{M}_D\nabla\tilde{B}$. This means that if we keep the strength of the real force (proportional to $\kappa e_D/m_V$) constant, the larger $m_V$ is, $\tilde{\bB}$ should be more sloped and the equivalent force on the guiding center should be stronger. Consequently, we expect that DM is more likely to be directly bounced away as if hitting a wall rather than wandering around in the solar interior in the case of strong dark magnetic force.

\section{Solar Reflected Dark Matter}
\label{sec:SRDM}
The plasma in the solar interior is composed of fast-moving electrons and ions, with temperatures ranging from 1 eV on the surface to 1 keV in the core. The solar temperature and density profiles~\cite{Bahcall:2004pz} are still debated, but for our purpose, the difference is negligible. Generally, the value of the scattering rate is governed by
\begin{equation}
    \Gamma_{\chi-i}=\frac{1}{2k_1^0}\int\frac{\dd^3 k_2}{2k_2^0(2\pi)^3}\int\frac{\dd^3 p_1}{2p_1^0(2\pi)^3}f_i(p_1)\int\frac{\dd^3 p_2}{2p_2^0(2\pi)^3}(2\pi)^4\delta^4(k_1+p_1-k_2-p_2)\sum_{\text{spin}}|\mathcal{M}|^2,
\end{equation}
where $i$ stands for the particle type and $f_i(p_1)$ is the thermal distribution
\begin{equation}
    f_i(p_1)=n_i(\bx)\left(\frac{2\pi}{m_i T(\bx)}\right)^{3/2}e^{-p_1^2/2m_i T(\bx)},
\end{equation}
in which $n_i(\bx)$ and $T(\bx)$ are the particle number density and the temperature at a given point $\bx$ in the Sun. $k_1$ and $k_2$ are the initial and final momenta of the DM particle, $p_1$ and $p_2$ are the initial and final momenta of charged particles in the Sun. $\mathcal{M}$ stands for the scattering amplitude, which is averaged over initial spins and summed over final spins. In this paper, we consider light DM with $m_\chi$ less than an MeV. In the numerical simulation, the interior of the Sun is divided into $\sim$2000 radial layers, each layer is accompanied by temperature and number density values, so that the scattering rate can be computed at any position where the DM particle is present.

The explicit form of the DM-electron(ion) scattering rate in the massless mediator case is calculated in Ref.~\cite{An:2021qdl} in detail. Here we only mention the crucial parts in the calculation. The amplitude of the scattering including the electron Debye screening effect is
\begin{equation}
    \mathcal{M}=\frac{\kappa e_D e\langle J^0_D J^0\rangle}{\varepsilon_L(q^2+m_V^2)},
\end{equation}
where the longitudinal permittivity is given by
\begin{align}
    \varepsilon_L(A)=&1+\frac{e^2 n_e}{Tq^2}\left[1-2A\int\frac{x\dd x}{(2\pi)^{1/2}}\tanh^{-1}(x/A)e^{-x^2/2}\right]\\
    =&1+\frac{e^2 n_e}{Tq^2}F_1(A),
\end{align}
where for the electron, $A_e=(q^0/q)(m_e/T)^{1/2}$. If the effect of ions with label $i$ is also considered, the expression changes to
\begin{equation}
    \varepsilon_L=1+\frac{e^2 n_e}{Tq^2}\left[F_1(A_e)+\sum_i\frac{Z_i^2 n_i}{n_e}F_1(A_i)\right],
\end{equation}
where $A_i=(q^0/q)(m_i/T)^{1/2}$. Following the derivation in Ref.~\cite{An:2021qdl}, the rate of DM-electron scattering can be written as
\begin{equation}
    \Gamma_{\chi-e}=\frac{n_e\kappa^2 e^2 e_D^2}{(2\pi T)^{3/2}m_e^{1/2}}\int\dd x_q\dd c_q\exp\left[-\frac{1}{2}\left(\frac{1}{2} x_q+A_e\right)^2\right]\frac{1}{|x_q^2+\frac{e^2 n_e}{Tq^2}F_1(A_e)|^2}\frac{x_q^5}{(x_q^2+\frac{m_V^2}{m_e T})^2},\label{gamma}
\end{equation}
where
\begin{equation}
    A_e=v_1\left(\frac{m_e}{T}\right)^{1/2}c_q+\frac{m_e}{2m_\chi}x_q,
\end{equation}
and $x_q=q/\sqrt{m_e T}$, $v_1=k_1/m_\chi$, $c_q=\cos\theta_{qk_1}$. In the simulation program, Eq.~\eqref{gamma} is used as a distribution to generate a random momentum transfer $q$ and an accompanied direction $\cos\theta_{qk_1}$. As mentioned in Ref.~\cite{An:2021qdl}, when $m_V\rightarrow 0$, the total rate of Coulomb-like scattering diverges. Even with the screening effect, the total rate is still too large for conducting simulations. However, divergent behavior is contributed mainly by forward scatterings, which have little effect on the DM energy or trajectory. For the purpose of SRDM, such possibilities should be cut off. Therefore, we introduce an IR cutoff represented by $\zeta$ to $x_q$:
\begin{equation}
    x_q=\frac{q}{\sqrt{m_e T}}=\frac{m_\chi \Delta v_\chi}{\sqrt{m_e T}}>x_{q,\text{cut}}=\frac{m_\chi\zeta v_{\chi0}}{\sqrt{m_e T}}=\zeta\frac{m_\chi}{m_e}v_{\chi0}\sqrt{\frac{m_e}{T}}.
\end{equation}
In this formula $v_{\chi 0}$ is a reference velocity, often taken as the average halo DM velocity $v_{\chi 0}=220\,\text{km}/\text{s}$. As shown in Ref.~\cite{An:2021qdl}, for scattering in the Sun, taking $\zeta=5$ is enough to capture the soft scattering behavior.

\begin{figure}
    \centering
    \includegraphics[width=0.6\linewidth]{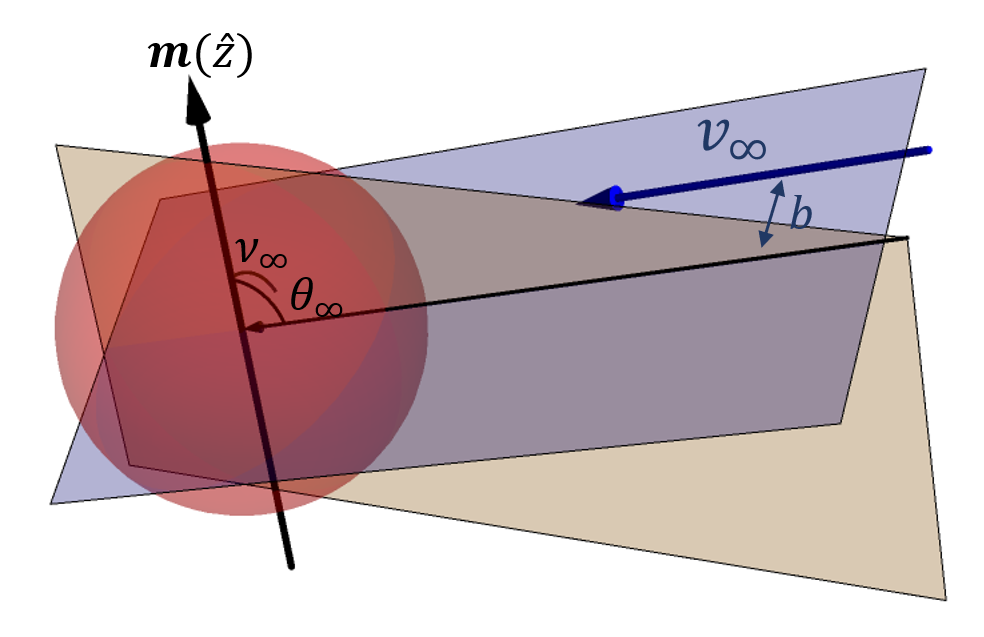}
    \caption{Geometrical illustration of initial parameters.}
    \label{fig:geo_illu}
\end{figure}

The numerical results of SRDM are derived through a Monte Carlo simulation program. The simulation follows similar steps as Ref.~\cite{An:2021qdl}, but is modified in order to account for the dark magnetic force:
\begin{enumerate}
    \item Randomly generate the initial impact parameter $b\in A_{\rho}$ ($A_\rho$ is the impact disc), the velocity at infinity $v_\infty$ and the angle of velocity according to the halo DM distribution; and then send the DM particle near the Sun ($r=4R_\odot$), taking into account energy and angular momentum conservations. The gravitational potential will accelerate the DM particle in this process. Since the solar dark magnetic field is axially symmetric, the angle between the velocity at infinity and solar magnetic axis, $\theta_\infty$, and the angle of aiming point on the impact disc, $\nu_\infty$, are also important. Therefore, $\theta_\infty$ and $\nu_\infty$ are also generated at random evenly. A geometrical illustration of these initial parameters is shown in Fig.~\ref{fig:geo_illu}.
    \item Draw the trajectory of the DM particle step by step. The magnitude of the velocity will change due to the gravitational force of the Sun. The main difference from the procedure in Ref.~\cite{An:2021qdl} is that, apart from gravity, the dark magnetic force [Eq.~\eqref{DLF}] is also considered. Because the distribution of the dark magnetic field is direction dependent and the force is perpendicular to the velocity, the 1D coordinate used in Ref.~\cite{An:2021qdl} is not appropriate and we are forced to use 3D Cartesian coordinates.
    \item At each step, if the DM particle is inside the Sun, generate a random number in the range $[0,1]$ and compare it to the scattering probability $P=\Gamma_{\chi-e}\Delta l/v$ (where $\Delta l$ is the step length and $v$ is the velocity of the DM particle) to determine whether a scattering happens or not. If a scattering happens, then use Eq.~\eqref{gamma} to generate $q$ and $\cos\theta_{q k_1}$. With the value and direction of the momentum transfer, the momentum of the DM particle is changed, sending the DM particle into a new trajectory with a new energy value. Return to step 2 until the DM particle is far enough from the Sun ($r>4R_\odot$).
    \item Collect the total energy $E_\chi$ of the DM particle at infinity, draw a histogram, and normalize it. Finally, we get the normalized reflected flux $F_{A_\rho}(E_\chi)$.
\end{enumerate}

\begin{figure}
    \centering
    \includegraphics[width=0.7\linewidth]{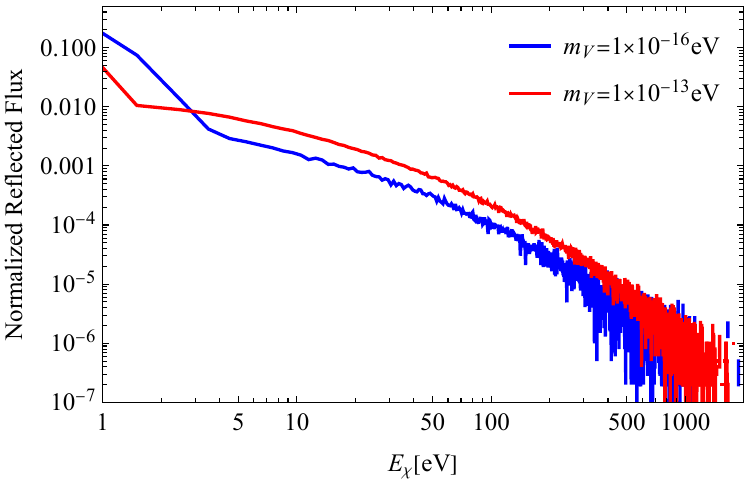}
    \caption{Normalized reflected flux with $m_\chi=0.5\,\text{MeV}$ and $Q_{\text{eff}}=\kappa e_D/e=1\times10^{-9}$. This flux only considers the DM population which comes close enough to the Sun to be reflected or gravitationally affected, and is normalized to 1. The blue curve has $m_V=1\times 10^{-16}\,\text{eV}$, while the red curve has $m_V=1\times 10^{-13}\,\text{eV}$. For $m_V=1\times 10^{-13}\,\text{eV}$, the strength of the dark magnetic field is so weak that the effect of the dark magnetic force is negligible, and the flux is the same as the flux without considering the dark magnetic field.}
    \label{fig:nflux}
\end{figure}

\begin{figure}
    \centering
    \includegraphics[width=0.7\linewidth]{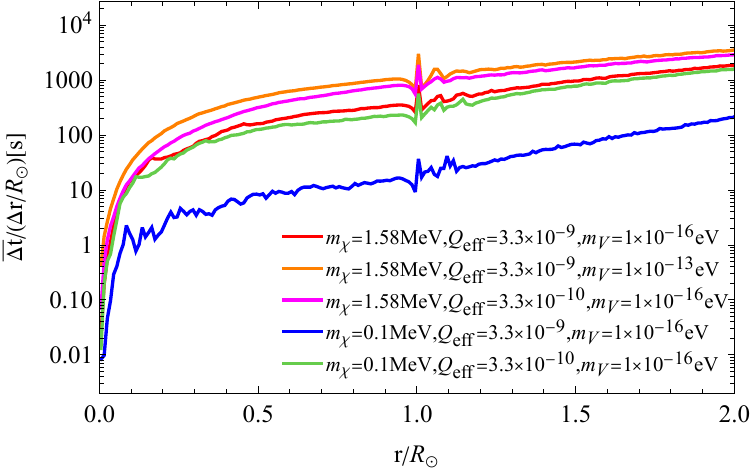}
    \caption{Event-averaged residence time $\bar{\Delta t}$ of the DM particles at each radius $r$ per radius interval $\Delta r$ normalized by solar radius $R_\odot$. Scatterings are not considered in this graph. The graph demonstrates that for larger dark magnetic force and smaller DM mass, the DM particle is more difficult to enter the core region of the Sun. Therefore, the high-energy tail of SRDM is suppressed by the presence of the dark magnetic field. The wiggling at $r/R_\odot=1$ is caused by the difference of $r$-stratification and the rules of step length in the interior and exterior of the Sun. When the DM particle crosses the surface of the Sun, the stratification is denser and the step length is smaller in the interior. Therefore, the recording of residence time may not be very accurate when the DM comes near the surface.}
    \label{fig:time_plot}
\end{figure}

After the normalized reflected flux spectrum is generated, the SRDM flux is related to the normalized reflected flux through
\begin{equation}
    \frac{\dd\Phi_{\text{SRDM}}}{\dd E_\chi}=\Phi_{\text{halo}}\times\frac{F_{A_\rho}(E_\chi)A_\rho}{4\pi r_\oplus^2},\label{eq:SRDM}
\end{equation}
where $r_\oplus\approx1\,\text{A.U.}$ is the distance between the Sun and Earth. An illustration of the reflected flux when $m_V=1\times 10^{-16}\,\text{eV}$ is shown in Fig.~\ref{fig:nflux}, compared to the case $m_V=1\times 10^{-13}\,\text{eV}$ where the effect of the dark magnetic field is negligible. This flux only accounts for the DM particles that come close enough to the Sun and are reflected or significantly gravitationally attracted by the Sun. The halo DM flux of the DM particles far away from the Sun is a separate population. As shown in Eq.~\eqref{eq:SRDM}, the reflected flux is suppressed by a geometric factor $A_\rho/4\pi r_\oplus^2\approx 8.65\times 10^{-5}$ compared to the halo DM flux, but has significantly higher velocities which can result in enhanced sensitivity in the detectors. The graph demonstrates that when the dark magnetic field is present, the reflected flux is enhanced around $E_R\sim 10\,\text{eV}$, while suppressed for $E_R>10\,\text{eV}$. This modification confirms the dual role of the dark magnetic field in SRDM physics. The trajectories of DM particles are redirected by the dark magnetic force inside the Sun, increasing their residence time and consequently the probability of more scatterings. At the same time, the hotter and denser core region is shielded by the dark magnetic field, suppressing keV-energy scatterings. Our simulations indicate that the shielding effect will eventually dominate, resulting in suppression of the keV-tail and enhancement of the $\sim 10\, \text{eV}$ spectrum. This core exclusion is quantified in ~\ref{fig:time_plot}, which plots the event-averaged residence time per unit radius. The distribution confirms that the net effect of the dark magnetic field generally reduces DM penetration into the solar core.

In Sec.~\ref{sec:solarDMF}, we discussed the generation of the solar magnetic field and emphasized that small-scale extreme values do not affect the qualitative estimates presented. However, given the possibility of very strong magnetic fields near the solar surface, a thin layer of dark magnetic field of the order of $\kappa\times$kG could, in principle, significantly alter the reflected flux. This raises a potential concern for the conclusions of this paper. First, we note that in Ref.~\cite{Velez:2008wu}, magnetic fields stronger than 200 G occupy only a small fraction of the solar photosphere. Moreover, these small-scale fields exhibit random orientations, leading to large-scale cancellation, and thus resulting in the SMMF value. Second, inferring the internal magnetic field of the Sun using helioseismology remains highly uncertain. As a result, a detailed modeling of the magnetic field structures discussed in Refs.~\cite{Velez:2008wu,HighMagneticField} is beyond the scope of this work.

\begin{figure}
    \centering
    \includegraphics[width=0.48\linewidth]{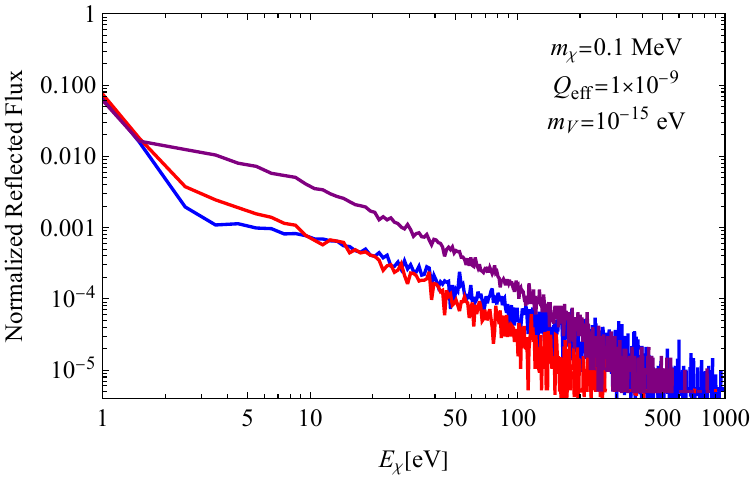}
    \includegraphics[width=0.48\linewidth]{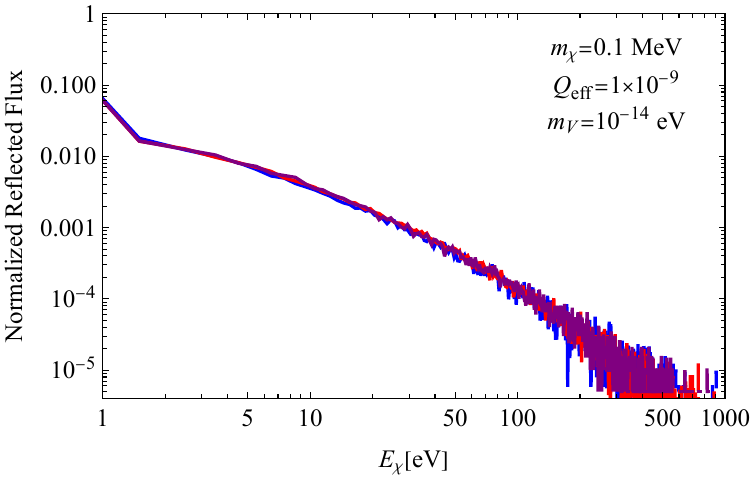}
    \caption{Normalized reflected flux of DM for different $m_V$. The blue curve shows the flux with the value of dark magnetic field magnified by a factor of 1000 for $0.996\,R_\odot<r< 1\,R_\odot$; The red curve shows the flux with the original model in Sec.~\ref{sec:solarDMF}; The purple curve shows the flux with the dark magnetic field turned off.}
    \label{fig:turb_compare}
\end{figure}

Nevertheless, to assess the potential impact of a thin layer of strong dark magnetic field, we artificially enhance the dark magnetic field strength by a factor of 1000 in the region $0.996\,R_\odot<r< 1\,R_\odot$. The simulation results are shown in Fig.~\ref{fig:turb_compare}. For small dark photon masses $m_V$ and small dark matter masses $m_\chi$ , the inclusion of this enhanced dark magnetic field layer increases the flux at high energies. However, the flux remains below the case without dark magnetic field at energies around 100--200 eV. A slight enhancement relative to the case without dark magnetic field is observed in the $\sim1\,\text{keV}$ tail, but the uncertainty is large in this energy range. Importantly, given the random orientation of the small-scale magnetic fields, this enhancement is likely an overestimate; the actual flux is expected to be lower. We therefore conclude that even under a great amplification of the dark magnetic field strength, the resulting modifications do not substantially affect our findings, and the qualitative conclusions of this paper remain valid. In the future, if helioseismology and direct observational data are combined to yield a concrete profile of the solar magnetic field, the potential role of thin layers of strong magnetic fields will become an important topic for further investigation.

\section{Direct Detection}
\label{sec:detection}
Finally, we can use the reflected flux to calculate the signal in ground-based direct detection experiments, such as XENONnT~\cite{XENON:2022ltv} and China Dark matter EXperiment (CDEX)-10~\cite{CDEX:2022kcd}. Generally, the averaged scattering cross section of the ionization process can be calculated by an integral over momentum transfer $q$~\cite{Essig:2011nj}:
\begin{equation}
    \frac{\dd\langle\sigma_e\rangle}{\dd\ln E_e}=\frac{\bar{\sigma}_e}{8\mu_e^2}\int\dd q\left[q|F_{\text{DM}}(q)|^2|f_e(p_e,q)|^2\eta(E_{\text{min}}(q,\Delta E))\right].
\end{equation}
Here, the recoil energy is $E_e=p_e^2/2m_e$, where $p_e$ is the outgoing electron momentum; $q$ and $\Delta E$ are the momentum and energy transfer; $\bar{\sigma}_{\chi e}=\kappa^2 e^2 e_D^2 \mu_e^2/\pi(q_0^2+m_V^2)^2$ is a reference cross section. The reference momentum transfer is often chosen to be $q_0=\alpha_{\text{EM}} m_e$. $|F_{\text{DM}}(q)|^2=1/q^2$ is the DM form factor. For our interest, $m_V<1\times 10^{-13}\,\text{eV}$, and the mass of the dark photon can always be ignored. $|f_e(p_e,q)|^2$ is the electron form factor that depends on the material and initial electron states. The $\eta$ function is the averaged squared inverse velocity,
\begin{equation}
    \eta(E_{\text{min}})=\int_{E_{\text{min}}}dE_\chi\frac{m_\chi}{2E_\chi}\frac{1}{\Phi_{\text{halo}}}\frac{d\Phi_{SR}}{dE_\chi},
\end{equation}
where
\begin{equation}
    E_{\text{min}}=\frac{1}{2}m_\chi v_{\text{min}}^2=\frac{1}{2}m_\chi\left|\frac{\Delta E}{q}+\frac{q}{2m_\chi}\right|^2.
\end{equation}
Therefore, the total ionization rate is
\begin{equation}
    \frac{dR_e}{d\ln E_e}=N_T\Phi_{\text{halo}}\frac{d\langle\sigma_e\rangle}{d\ln E_e},
\end{equation}
where $N_T$ is the number of atoms in the target.

\begin{figure}
    \centering
    \includegraphics[height=0.315\linewidth]{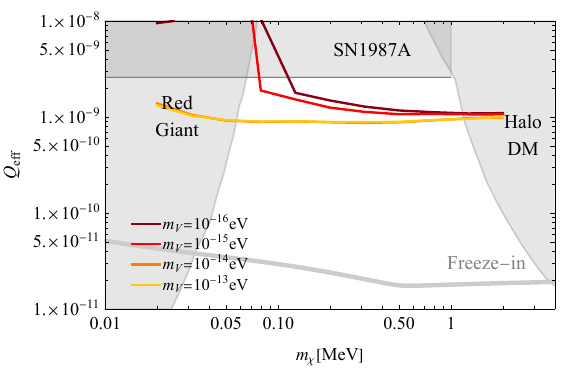}
    \includegraphics[height=0.315\linewidth]{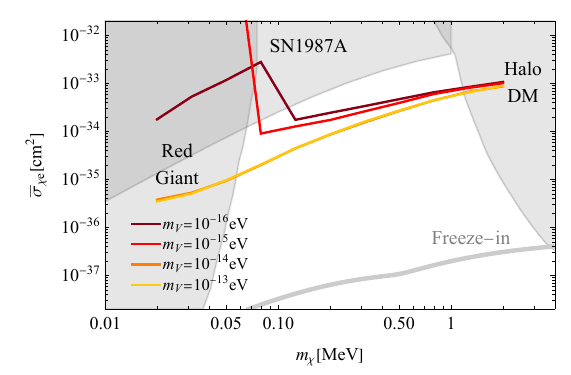}
    \includegraphics[height=0.315\linewidth]{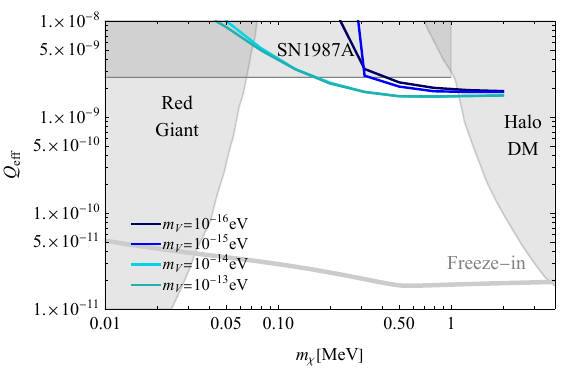}
    \includegraphics[height=0.315\linewidth]{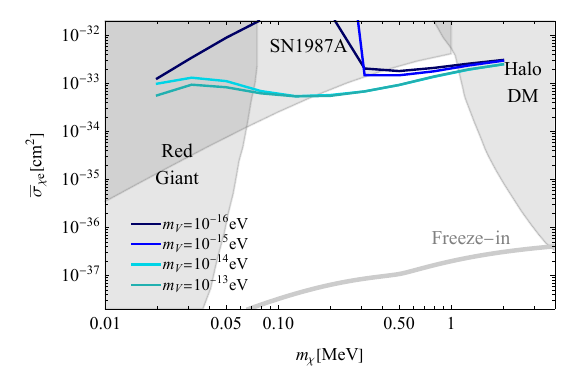}
    \caption{Constraints on DM millicharges $Q_{\text{eff}}=\kappa e_D/e$ and reference cross sections $\bar{\sigma}_{\chi e}$ for different dark photon masses $m_V$. The first row shows the constraints from CDEX-10~\cite{CDEX:2022kcd}, and the second row shows the constraints of XENONnT~\cite{XENON:2022ltv}. The astrophysical cooling constraints are also shown as a comparison. The shaded regions on the left are the red giants cooling constraints from Ref.~\cite{Fung:2023euv}, the shaded regions on the right are a combination of various halo DM direct detection constraints from Refs.~\cite{Essig:2017kqs,SENSEI:2020dpa,SENSEI:2023zdf,DAMIC:2019dcn,DAMIC-M:2025luv}, the shaded regions at the top are the constraint from the cooling of supernova SN1987A~\cite{Fiorillo:2024upk}. At the bottom lies the freeze-in benchmark line~\cite{Essig:2011nj,Essig:2015cda,Chu:2011be,Dvorkin:2019zdi,Emken:2024nox,Bhattiprolu:2023akk,Bhattiprolu:2024dmh}.}
    \label{fig:scanplot}
\end{figure}

In the XENON experiment, electron recoils in the xenon target produce ionization signals (S2). For recoil energies above $1\,\text{keV}$ ($E_e \gtrsim 1\,\text{keVee}$), they additionally produce scintillation signals (S1). In this paper, we employ the XENONnT S1+S2 data from Ref.~\cite{XENON:2022ltv}. To obtain the xenon ionization rate, we calculate the ionization form factors $|f_{nl}(p_e,q)|^2$ for atomic orbitals labeled $nl$ following the procedure in Ref.~\cite{Essig:2017kqs} using the atomic wave functions from Ref.~\cite{Bunge1993RHF}. We split the 0--10 keV energy range into 10 bins. The expected event count in each bin $i$ for exposure time $T$ is
\begin{equation}
    N_i=T\int_{\ln E_{e,\text{min},i}}^{\ln E_{e,\text{max},i}} d\ln E_e\,\text{Eff}(E_e)\frac{dR_e}{d\ln E_e},
\end{equation}
where $\text{Eff}(E_e)$ is the efficiency function from Ref.~\cite{XENON:2022ltv}, $E_{e,\text{min},i}$ and $E_{e,\text{max},i}$ are the lower and upper bounds of each energy bin, respectively. After obtaining $N_i$, accompanied by measured counts $S_i$ and background estimates $B_i$ taken from Ref.~\cite{XENON:2022ltv}, we draw the constraint line in a simplified and conservative method. The signal event number in each bin $S_i$ is treated as a Poisson random variable with expectation values $\mu_i=N_i+B_i$ and variances $\sigma_i^2=N_i+B_i$. If $N_i+B_i\gg 1$, they can be approximated by Gaussian random variables with the same parameters. Therefore, we can define a logarithmic likelihood ratio:
\begin{equation}
    -2\ln\lambda=\sum_{i=1}^{10}\left[-\ln\left(\frac{S_i}{N_i+B_i}\right)+\frac{(S_i-N_i-B_i)^2}{N_i+B_i}\right].
\end{equation}
If $N_i+B_i$ in any bin is around or smaller than 1, the Gaussian approximation breaks down, and we have to replace the $-2\ln L$ of likelihood $L$ by
\begin{equation}
    \left[2(N_i+B_i)-2S_i\ln(N_i+B_i)+2\ln(S_i !)\right]
\end{equation}
in this bin. For large samples, this ratio follows a $\chi^2(10)$ distribution. If $-2\ln\lambda>18.307$, we reject the DM hypothesis with 95\% confidence, excluding the parameter point. The final constraints on the DM millicharge $Q_{\text{eff}}=\kappa e_D/e$ are shown in the right graph of Fig.~\ref{fig:scanplot}.

The CDEX experiment uses single-crystal semiconducting germanium as the target material, where the event rate is the sum of Bragg scattering of inner-shell tightly bound electrons and valence shell scattering of crystal Bloch state electrons. The calculations of the form factors follow the method in Ref.~\cite{An:2025cik}. Inner-shell scattering is similar to the xenon case and uses atomic ionization form factors $|f_{nl}(p_e,q)|^2$ derived from atomic orbitals in Ref.~\cite{Bunge1993RHF}. As germanium is a crystal with a lined-up lattice structure, interference between the scatterings on each atom causes coherent Bragg scatterings with discrete momentum transfers~\cite{An:2025cik}. This is crucial for $m_\chi\lesssim 0.05$ MeV, and the small-mass part of the constraint line from Ref.~\cite{CDEX:2023wfz} is corrected in the calculation of this paper. Valence electron form factors use Bloch wave functions computed with \texttt{Quantum ESPRESSO}~\cite{Giannozzi2017AdvancedCF} and lattice parameters from Ref.~\cite{germanium}.

With CDEX-10 signal and background data from Ref.~\cite{CDEX:2018lau}, we integrate event rates between the threshold 160 eV and 260 eV:
\begin{equation}
    N=T\int_{160\,\text{eV}}^{260\,\text{eV}}dE_e\left(\frac{dR_{\text{in}}}{dE_e}+\frac{dR_{\text{val}}}{dE_e}\right).
\end{equation}
With measured counts represented by $S$ and background by $B$, we can analyze constraints by treating $S$ as a Poisson random variable with expectation $\mu=N+B$ and variance $\sigma^2=N+B$. For $N+B\gg 1$, we can also approximate it by a Gaussian random variable with the same parameters. The logarithmic likelihood ratio is constructed as follows:
\begin{equation}
    -2\ln\lambda=-\ln\left(\frac{S}{N+B}\right)+\frac{(S-N-B)^2}{N+B}.
\end{equation}
This statistic approximates a $\chi^2(1)$-distributed random variable for large samples. We reject the DM model with a confidence level of 95\% if $-2\ln\lambda>3.841$. Finally, the constraints on DM millicharge are shown in the left graph in Fig.~\ref{fig:scanplot}.

As discussed in Sec.~\ref{sec:eom}, the dark magnetic force $F_{\text{dark}}\propto Q_{\text{eff}}/m_\chi m_V^2$. Therefore, the deflection effect is stronger for smaller $m_V$ and smaller $m_\chi$. Especially for $m_V\lesssim 10^{-15}\text{eV}$ and $m_\chi\lesssim 0.1\,\text{MeV}$, the deflection becomes so strong that the DM is severely prevented from entering the core region of the Sun, resulting in a great weakening of the reflected DM constraints. For $m_V\gtrsim10^{-14}\text{eV}$, the solar dark magnetic field is too weak to affect the motion of DM.

\section{Summary}
\label{sec:summary}
Traditionally, the mass of the dark photon is solely treated as a parameter in the propagator of the DM scattering amplitude. In this paper, we demonstrate that dark photons with very small mass can potentially cause astrophysical-scale phenomena simultaneously. Due to the small mixing between the dark photon and the normal photon, any current coupled to the normal electromagnetic field will also source a dark photon field. The resulting dark magnetic field follows the shape of the Yukawa potential, decaying with a length scale corresponding to $m_V^{-1}$. Therefore, if $m_V$ is comparable to an astronomical scale (the solar radius), corresponding to $m_V\sim 10^{-15}\text{eV}$, then the dark magnetic field generated by the internal dynamics of the Sun is sufficiently strong to alter the trajectories of DM particles approaching the Sun. The effect of the dark magnetic field exhibits a double nature: on the one hand, it will bend the trajectories of DM particles, forcing DM particles to travel along longer paths inside the Sun, potentially enhancing opportunities for high-energy scatterings. On the other hand, if the incident DM particle has an unfavorable initial pitch angle relative to the dark magnetic field, then the DM particle has a large probability of being deflected away and prevented from entering the hotter core region of the Sun. Consequently, this will suppress useful keV-energy DM-electron scatterings. In order to find out the overall effect, we developed a Monte Carlo simulation program based on pre-existing works on SRDM, and incorporated new ingredients to capture the effect of the dark magnetic field. It turns out that the core-shielding effect dominates, and the high energy tail of SRDM is always suppressed compared to the case without a dark magnetic field in the parameter space region we are interested in. Consequently, the sensitivity of ground-based experiments such as CDEX-10 and XENONnT is significantly weakened for $m_V\lesssim 10^{-15}\text{eV}$ and $m_\chi\lesssim 0.1\,\text{MeV}$. Although the solar dynamics and the distribution of the dark magnetic field are greatly simplified in this paper, the main point is still enlightening in the sense that the global configuration of the dark photon field is as important as its excitations. Future work should integrate realistic solar magnetohydrodynamics--including transient features like sunspots, flares, and other strong field regions--to refine predictions of SRDM flux and its implications for direct detection.


\acknowledgments

The author would like to thank PhD supervisor Haipeng An for instructions and helpful discussions. This work is supported in part by the National Key R\&D Program of China under Grants No. 2021YFC2203100 and No. 2017YFA0402204, the NSFC under Grants No. 12475107 and No. 12525506.


\section*{Data availability}
The data that support the findings of this article are not publicly available upon publication because it is not technically feasible and/or the cost of preparing, depositing, and hosting the data would be prohibitive within the terms of this research project. The data are available from the authors upon reasonable request.

\bibliography{ref.bib}

\end{document}